\documentclass[aps,pre,twocolumn,superscriptaddress]{revtex4}

\usepackage{t1enc}
\usepackage{amsmath}
\usepackage{graphicx}
\usepackage{epsfig}
\usepackage{psfrag}
\usepackage{times}

\begin{document}

\title{Dynamics of systems with isotropic competing interactions in an external field:
a Langevin approach}

\author{Rogelio D\'\i az-M\'endez}
\affiliation{Nanophysics Group, Department of Physics,
Electric Engineering Faculty, %\\
CUJAE, Ave 114 final, La Habana, Cuba}
\affiliation{``Henri-Poincar\'e-Group'' of Complex Systems, Physics Faculty, University of
Havana, La Habana, CP 10400, Cuba}

\author{Alejandro Mendoza-Coto}
\affiliation{Department of Theoretical Physics, Physics Faculty, University of
Havana, La Habana, CP 10400, Cuba}
\affiliation{``Henri-Poincar\'e-Group'' of Complex Systems, Physics Faculty, University of
Havana, La Habana, CP 10400, Cuba}

\author{Roberto Mulet}
\affiliation{Department of Theoretical Physics, Physics Faculty, University of
Havana, La Habana, CP 10400, Cuba}
 \affiliation{``Henri-Poincar\'e-Group'' of Complex Systems, Physics Faculty, University of
Havana, La Habana, CP 10400, Cuba}

\author{Lucas Nicolao}
\affiliation{Dipartimento di Fisica, Universit\`a di Roma "La Sapienza", P.le Aldo Moro 2, 00185 Roma, Italy}

\author{Daniel A. Stariolo}
\affiliation{Departamento de F\'{\i}sica, Universidade Federal do Rio Grande do Sul and\\
National Institute of Science and Technology for Complex Systems\\ CP 15051, 91501-970, Porto Alegre, Brasil}

\date{November 2010}

\begin{abstract}
We study the Langevin dynamics of a ferromagnetic Ginzburg-Landau
Hamiltonian with a competing long-range repulsive term in the presence
of an external magnetic field. The model is analytically solved within
the self consistent Hartree approximation for two different initial
conditions: disordered or {\em zero field
cooled} (ZFC), and fully magnetized or {\em field
cooled} (FC). To test the predictions of the approximation we develop
a suitable numerical scheme to ensure the isotropic nature of the
interactions. Both the analytical approach and the numerical simulations of
two-dimensional finite systems confirm a simple aging scenario at zero
temperature and zero field. At zero temperature a critical field
$h_c$ is found below which the initial conditions are relevant
for the long time dynamics of the system. For $h < h_c$ a logarithmic
growth of modulated domains is found in the numerical simulations but
this behavior is not captured by the analytical approach which predicts
a $t^{1/2}$ growth law at $T=0$.
\end{abstract}

\maketitle

%****************************************************************************
\section{Introduction}

It is now well understood that modulated structures can arise from
underlying competing interactions acting upon different scales
\cite{SeAn1995}. In magnetic systems, e.g. in ultrathin magnetic
films with perpendicular anisotropy \cite{PoVaPe2003, DeMaWh2000},
these phases are present due to an exchange (ferromagnetic) short
range interaction competing with a weaker but long range
(antiferromagnetic) dipolar interaction.  This competition gives rise
to striped and bubbled patterns that break the translational and/or
rotational symmetry of the system.

This phenomenology is shared with a large number of systems. Type-II
superconductors \cite{TrStAxNaUc1995}, doped Mott insulators
\cite{KiFrEm1998}, quantum Hall systems \cite{BaFrKiOg2002} and
diblock copolymers \cite{FrBi1989,FrBa1996} are other examples of
systems with competing interactions which have symmetry breaking
transitions to inhomogeneous phases. Since domains in these
systems often are of mesoscopic scales, it is possible to
perform a general analysis which neglects the microscopic details
of each specific system while capturing its general properties. A
pioneering attempt in this direction was done by Brazovskii
\cite{Br1975}, who showed that systems in which the
spectrum of fluctuations attain a minimum in a shell 
at a non-zero wave vector, undergo a first order phase transition
to a modulated state driven by fluctuations, 
in contrast to the second order transition
predicted by the mean field theory.
Moreover, the presence of competing interactions frequently lead to a
proliferation of metastable states at low temperatures that may
dominate the dynamical behavior of the system. For the
three dimensional case, recent experimental results in colloidal
systems \cite{KlRoTa2010} report the presence of glassy states,
in agreement with theoretical predictions for a model with screened
electrostatic repulsion \cite{TaCo2006} and for a 
Coulomb frustrated ferromagnet \cite{WeScWo2001,grousson01,grousson02}.

However, it is still far from clear which are the relevant aspects
that control the long time dynamics of these systems and, more generally, 
what kind of dynamics should one expect. In an early work for a model with
ferromagnetic and dipolar interactions, Roland and Desai \cite{roland90}
studied the dynamics in the short-time regime, showing that
the order parameter evolves to modulated structures. Almost
contemporaneously, Elder and Grant \cite{ElGr1990}
%, through a perturbation treatment 
obtained a scaling relation for the dynamics of phase
ordering of the modulated phases, characterized by a growth law $L\propto
t^{1/z}$ with a dynamical
exponent $z=2$. After that, many works were devoted to verify such 
scaling relation, solving numerically the dynamics of the Swift-Hohenberg 
model \cite{SwHo1977,HoSw1995}.
Indeed most of these numerical simulations \cite{ElViGr1992, CoMe1995,
HouSasa1997, ChBr1998, QiMa2003} report a late-time regime that
satisfies scaling, but with a growth exponent near $z=5$ for
zero temperature quenches and $z=4$ for low finite temperatures,
obtained mainly monitoring the structure factor and related quantities. 
%Worth mentioning that in some of these works the exponents estimated
%were slightly different for other observing quantities, as in the
%evolution of energy and domain wall lenght \cite{HouSasa1997} and
%orientational correlation lenght \cite{ChBr1998} scaling as $z=1/4$
%and $1/3$ for zero and low temperatures, respectively - 
%In \cite{QiMa2003} the authors report $z=1/3$ at zero temperature
%from measurments of energy, orientational correlation lenght and
%defects sizes and densities. An exception is reported in
In \cite{BoVi2001} an exponent $z=3$ was observed in zero temperature 
quenches which was associated to grain boundary motion.
For quenches at zero temperature, a dynamic crossover to a frozen state 
was reported \cite{SaDe1994}, and associated with grain
boundary pinning \cite{BoVi2002}. On the other hand, recent
experiments in diblock copolymers \cite{harrison2002}(that are assumed 
to be good model systems for studying phase ordering of modulated
structures), show that the late time dynamics of the
stripe ordering satisfies scaling with $z=4$, driven mainly by the annihilation involving
tripoles and quadrupoles of disclinations. For a bubble forming system
 Harrison et al. \cite{harrison2004} found the same growth
law, but this time dominated by the collapse of smaller
grains which reside on the boundary of two larger
grains. More recently, Gomez et al. \cite{GoVaVe2006} found a logarithmic growth
$L(t) \simeq \ln{t}$ of domains in simulations of an hexagonal diblock
copolymer model. The logarithmic growth is attributed to the Lifshitz
mechanism of pinning of triple points at grain boundaries. Here we report
logarithmic growth of different characteristic length scales in simulations of a
stripe forming system. Also, Gleiser et al. \cite{GlTaCaMo2003} found a
crossover between an apparently logarithmic relaxation to a $t^{1/2}$ law
in the coarsening dynamics of a dipolar frustrated Ising model. 
So, despite a lot of numerical and experimental effort the question of
dynamical universality of modulated systems remains elusive.

Recently, Mulet and Stariolo \cite{mulet07} solved the long time Langevin 
dynamics of a model with competing interactions in the Hartree (self-consistent field)
approximation. For the case of zero external field they
showed that the dynamics can be separated in two times scales. In the
first one modulated structures emerge, and the behavior resembles
the formation of typical ferromagnetic domains \cite{NeBr1990}. Once
these modulated structures are formed the dynamics changes
qualitatively, becoming independent of the system dimension and 
the temperature is a relevant variable. For $T>0$ the system exhibits
interrupted aging and a standard paramagnetic (exponential) relaxation for large
times. At low temperatures domains of stripes of finite size, characterized
by a finite correlation length, are formed. At $T=0$  
the correlation length diverges and stripes order sets in with a
growth law $L\propto t^{1/2}$. Our simulations show that these results are
correct 
%in the high temperature phase or, at very low temperatures 
in the time scale while the very slow dynamics of topological defects 
does not play a significant role. 

In this work we extend the results of \cite{mulet07} by solving,
within the Hartree approximation, the same model in the presence of an
external magnetic field and for two different initial
conditions. The initial conditions were chosen to mimic standard
experimental protocols: in the {\em Zero Field Cooling} (ZFC)
experiment, the system is initialized in a highly disordered state (high
temperature and zero external field). After some waiting time the temperature 
is suddenly lowered towards the modulated phase and the field is turned on. 
In the {\em Field Cooling} (FC) protocol, the sample is first fully
magnetized in the presence of an external field, and then the
temperature is decreased keeping the field on. To check the
validity of our approximations, we develop a reliable numerical scheme
to reach the asymptotic dynamical behavior, while ensuring the isotropic nature
of the model interactions. We compare the analytical results
with extensive numerical simulations of a two-dimensional magnetic system 
with dipolar interactions.

The rest of the paper is organized as follows. In section \ref{aa} we
present the model and the Hartree
approximation to solve its dynamics. We present analytical results for
the autocorrelations and spatial correlation functions for 
zero temperature ($T=0$) and finite external field ($H$).  Section \ref{ns} is
devoted to the numerical simulations. It begins with an analysis of
the fluctuation spectrum of the model, which is essential to 
compare the simulations of finite systems and the
analytical results. Then we present the numerical scheme used
and compare the simulations with the analytical
predictions. Finally in section \ref{conc} we summarize and present
the conclusions of our work. The details of the calculations are left
for the appendices.

%*****************************************************************************
\section{Analytical approach}
\label{aa}

We consider an effective Ginzburg-Landau Hamiltonian :

\begin{eqnarray}
\nonumber
 {\cal H}[\phi]  &=& \int d^d x \left[ \frac{1}{2} (\nabla \phi(\vec{x}))^2
+\frac{r}{2}\phi^2(\vec{x}) \right.\\ \nonumber
&+&\left. \frac{u}{4} \phi^4(\vec{x}) - h(\vec{x})\phi(\vec{x}) \right] \\ 
&+& \frac{1}{2\delta} \int d^d x\,d^d x'\
\phi(\vec{x})\,J(\vec{x},\vec{x}')\,\phi(\vec{x}')
\label{hamilt}
\end{eqnarray}

\noindent where $r<0$, $u>0$ and $J(\vec{x},\vec{x}') =
J(|\vec{x}-\vec{x}'|)$ represents a repulsive, isotropic, competing
interaction. $h(\vec{x})$ is an external field and the
parameter $\delta$ measures the relative intensity between the
attractive and repulsive parts of the Hamiltonian. In the limit
$\delta \to \infty$ one recovers the ferromagnetic $O(N)$ model (for
$N=1$) \cite{CoLiZa2002, NeBr1990, ChCuYo2005}.

\noindent Following the general formalism presented in
Ref. \cite{mulet07}, the Langevin dynamics for Hamiltonian (\ref{hamilt}) takes the form:
\begin{eqnarray}
\nonumber
\frac{1}{\Gamma}\frac{\partial \phi(\vec{x},t)}{\partial t}&=& \nabla^2 \phi(\vec{x},t) -r \phi(\vec{x},t)-
u\,\phi^3(x,t)  + h(\vec{x},t)\\ \label{geq}
&-& \frac{1}{\delta}\int d^d x'\,J(\vec{x},\vec{x}')\phi(\vec{x}',t)+\frac{1}{\Gamma}\eta(\vec{x},t)
\label{eqts}
\end{eqnarray}
%where $\Gamma$ is set to $1$ for simplicity and 
where $\eta(\vec{x},t)$ represents a Gaussian noise that 
models the coupling to a heat bath.

Within the self-consistent (Hartree) approximation the non-linear term $\phi^3$ is
substituted by \( 3\langle \phi^2(\vec{x},t)\rangle \phi(\vec{x},t) \)
where the average is performed over the initial conditions and noise
realizations. It is worth to note that, within the approximation,
$\left\langle\phi^2(\vec{x},t)\right\rangle$ is spatially homogeneous,
which implies a restriction in the possible ensembles of initial
conditions taken to perform the averages.

We have studied two kinds of initial conditions: 
disordered (Zero Field Cooled) and ferromagnetic (Field Cooled). Both 
can be written in the general form
\begin{eqnarray}
\nonumber
\langle\phi(\vec{x},0)\rangle&=&\phi_0 \\
\langle\phi(\vec{x},0)\phi(\vec{x}',0)\rangle&=&v(\vec{x}-\vec{x}')
\label{gcond}
\end{eqnarray}

In this way, the disordered condition ($h=0$ and $T >> 0$) is obtained
with $\phi_0=0$ and
$v(\vec{x}-\vec{x}')=\Delta\delta(\vec{x}-\vec{x}')$, while the
ferromagnetic initial condition  ($h>>0$) is represented by $\phi_0>0$ and
$v(\vec{x}-\vec{x}')=\phi_0^2$.

After applying Fourier transforms, the equation of motion (\ref{geq}) in
the Hartree approximation reads:
\begin{equation}
\frac{\partial \hat{\phi}(\vec{k},t)}{\partial t}= -[A(k)+I(t)]\ \hat{\phi}(\vec{k},t) + 
\hat{\eta}(\vec{k},t)+\hat{h}(\vec{k},t),
\label{lan}
\end{equation}
with
\begin{equation}
I(t) = r_0 + g\, \langle \phi^2(\vec{x},t)\rangle,
\label{I}
\end{equation}
\begin{equation}
A(k)=k^2+ \frac{1}{\delta}\, \hat{J}(k) - a_0,
\label{AK}
\end{equation}
where $g=3u$,
 $a_0=k_0^2+ 1/\delta \ \hat{J}(k_0)$ and $r_0=r+a_0$.  $k_0$ is the
wave vector minimizing the spectrum of fluctuations $A(k)$.
In the case of isotropic competing interactions
it is well established that this quantity has a minimum on a
spherical shell of radius $k_0$ in reciprocal space \cite{Br1975}. Thus, 
definitions (\ref{I}) and (\ref{AK}) ensure that $A(k_0)=0$.

The general solution of the dynamical equations (\ref{lan}) can be written as:

\begin{eqnarray}
\nonumber
\hat{\phi}(\vec{k},t)&=&\hat{\phi}(\vec{k},0)\,R(\vec{k},t,0)+\int_0^t R(\vec{k},t,t')\,
\hat{\eta}(\vec{k},t')\,dt'  \\ \nonumber 
&&+ \int_0^t R(\vec{k},t,t')\,\hat{h}(\vec{k},t')\,dt'
\end{eqnarray}
where
$
R(\vec{k},t,t')=\frac{Y(t')}{Y(t)}e^{-A(k)(t-t')}
$
is the response function and
$
Y(t)=e^{\int_0^t dt' I(t')}
$.
The Gaussian thermal noise $\eta(\vec{x},t)$ satisfies
\begin{eqnarray}
\nonumber
\langle \hat{\eta}(\vec{k},t) \rangle & =  & 0, \\
\langle \hat{\eta}(\vec{k},t) \hat{\eta}(\vec{k}',t') \rangle & = & 2 \Gamma T(2\pi)^d
\delta(\vec{k}+\vec{k}') \delta (t-t')
\label{noise}
\end{eqnarray}
and a stationary homogeneous external field $(h(\vec{x},t)=h)$ is defined by:
\begin{eqnarray}
\nonumber
\langle \hat{h}(\vec{k}) \rangle & = &h(2\pi)^d \, \delta(\vec{k}),\\
\langle \hat{h}(\vec{k})\ \hat{h}(\vec{k}') \rangle & = &h^2 (2\pi)^{2d}\, \delta(\vec{k})\,\delta(\vec{k}') .
\label{field}
\end{eqnarray}

A complete solution amounts to the determination of the response function, or equivalently, $Y(t)$. 
Defining $K(t) = Y^2(t)$ and following standard procedures \cite{mulet07}, it is possible to reduce 
the problem to the solution of the following differential equation:
\begin{eqnarray}
\nonumber
\frac{dK(t)}{dt}&=&2r_0K(t)+2gV(t)\\ \nonumber 
&+&4gT\int_0^t d\tau \ K(\tau) f(t-\tau) \\ \nonumber
&+&4gh \ K(t)^{1/2}\phi_0 e^{-A(0)t}\ \times \\ \nonumber
&\ &\ \times \int_0^td\tau\ K(\tau)^{1/2}e^{-A(0)(t-\tau)}\\ 
&+&2gh^2\left[\int_0^t d\tau\ K(\tau)^{1/2}e^{-A(0)(t-\tau)} \right]^2 
\label{K}
\end{eqnarray}

where 
\begin{equation}
V(t) = \frac{1}{(2\pi)^d}\int d^dk\ e^{-2A(k)t}\, \hat{v}(\vec{k})
\label{Gt}
\end{equation}

and

\begin{equation}
f(t) = \frac{1}{(2\pi)^d}\int d^dk\ e^{-2A(k)t}.
\label{ft}
\end{equation}

$V(t)$ contains the information about the initial conditions, and the
last two terms in (\ref{K}) reflect the presence of the external
field.  This external field is associated to strongly non-linear
contributions in $K(t)$ that prevents us from using a standard Laplace
transformation \cite{mulet07} to solve equation (\ref{K}).
To deal with these non-linearities we propose an
appropriate ansatz for the long time behavior of $K(t)$ and then,
keeping consistency, follow usual techniques for solving
differential equations. We introduce the ansatz:

\begin{equation}
\int_0^t K^{1/2}(t')\,e^{-A(0)(t-t')}dt' = \zeta \, K^{1/2}(t) 
\label{hyp1}
\end{equation}

\noindent and solve for $K(t)$, determining $\zeta$
self-consistently. With this ansatz, the problem has been reduced to
find a solution for a linear differential equation in $K(t)$, a
function that encloses conditions (\ref{gcond}), (\ref{noise}) and
(\ref{field}). Once $K(t)$ is known, observables like response
functions, two times autocorrelations, and spatial correlation
functions can be readily obtained. 
Spatial correlations are given by:
\begin{eqnarray}
\nonumber
&&C(\vec{x},\vec{x}',t,t)=C(R,t)=\int \frac{d^dk}{(2\pi)^d}\times\\
&&\times\int
\frac{d^dk'}{(2\pi)^d}\,e^{i(\vec{k}\cdot\vec{x}+\vec{k}'\cdot\vec{x}')}\,
\langle\hat{\phi}(\vec{k},t)\,\hat{\phi}(\vec{k}',t)\rangle
\end{eqnarray}
where $R=\vert\vec{x}-\vec{x}'\vert$, while two-times autocorrelations are
defined as:
\begin{eqnarray}
\nonumber
&&C(\vec{x},\vec{x},t,t')=C(t,t')=\int \frac{d^dk}{(2\pi)^d}\times\\ 
&&\times\int
\frac{d^dk'}{(2\pi)^d}\,e^{i(\vec{k}+\vec{k}')\cdot\vec{x}}\,
\langle\hat{\phi}(\vec{k},t)\,\hat{\phi}(\vec{k}',t')\rangle.
\end{eqnarray}

Below we
present and discuss the analytical predictions for some of these
observables and leave to Appendix \ref{Appendix-A} the 
details of the calculations.

%**************************************************************************************
\subsection{Observables at $T=0$}
\label{ot0}

At zero temperature and in both the disordered (zero field cooled) and ferromagnetic
(field cooled)
quenches, we have found a critical field which divides the analysis in three regions
$h<h_c$, $h=h_c$ and $h>h_c$, with qualitatively different long time dynamics (
the critical filed is defined in Appendix \ref{Appendix-A}).
Starting from the ferromagnetic
configuration we show that, despite the existence of this critical
field, the decay of the correlation functions are essentially
exponential, although with different exponential or constant
rates. However, if the quench is done starting from a disordered high
temperature phase, we find important functionality changes in the
two-time and the spatial correlation functions as a function of $h$.

%***********************************
\subsubsection{Zero Field Cooling }

In table \ref{tab1} we show the predicted long time behavior of the spatial and two-times
autocorrelation functions in ZFC experiments.  

\begin{table}[ht]
\begin{center}
\begin{tabular}{|c|c|c|}
\hline
$h$  & $C(t,t')$ & $C(R,t)$ \\
\hline
$h<h_c$ & $C_{\infty}+\frac{(tt')^{\frac{1}{4}}}{(t+t')^{\frac{1}{2}}}$ & $f(R)+C_{\infty}$ \\
\hline
$h=h_c$ & $C_{\infty}+\frac{1}{(tt')^{\frac{1}{4}} (t+t')^{\frac{1}{2}}}$ & $f(R)/t+ C_{\infty}$\\
\hline
$h>h_c$ &  $C_{\infty}+\frac{e^{-\frac{1}{2}\lambda(t+t')}}{(t+t')^{\frac{1}{2}}} $ & $f(R)e^{-\lambda t}/\sqrt{t}+ C_{\infty}$\\
\hline
\end{tabular}
\caption{Long time solutions for $C(t,t')$ and $C(R,t)$ at $T=0$ for different values of the 
applied field in ZFC case.}
\label{tab1}
\end{center}
\end{table}

$C_{\infty}$ is a constant which is different in each expression and 
\begin{equation}
f(R)=\frac{1}{(k_0R)^{\frac{d}{2}-1}}\,J_{{\frac{d}{2}-1}}(k_0R)\,e^{-\frac{R^2}{4A_2t}},
\label{growthlaw}
\end{equation}
where $J_n(x)$ is a Bessel function of the first kind.

As can be seen, for $h\leq h_c$ autocorrelations show aging behavior.
In particular, the asymptotic ($t \gg t'$) two-times correlation function has
the form $C(t,t')\propto(L(t)/L(t'))^\nu$ that characterizes a
coarsening dynamics caused by the competition of domains of modulated
structures~\cite{mulet07}.
The typical size of these domains grows as $L(t)\propto t^{1/2}$, as
can be confirmed from the exponential term of $f(R)$ in the $h<h_c$
spatial correlations.  On the other hand, for $h > h_c$ the autocorrelations
 relax exponentially fast to the asymptotic value .
%This behavior is typical of models with
%non-conserved order parameter and ensures that the autocorrelation
%exponent \cite{FiHu1988} is $\lambda=1/2$. 
Also, it is worth to note
the independence of the two-times autocorrelations with the 
dimensionality of the system.

The spatial correlations at zero temperature show that only for
$h<h_c$ the system establishes modulated structures at infinite
time.

%************************************
\subsubsection{Field Cooling}

Because the initial condition is homogeneous ($\phi(\vec x)$ is constant),
 the correlations
in this case are independent of $R$. 

As already mentioned, in this case  
$K(t)$ is essentially exponential for any value of $h$ ( see
Appendix \ref{skt} for more details). Moreover, the dominant
exponential terms depend strongly on $h$. In particular, for $h=0$
one finds that the form of $C(t,t')$ depends on the ratio between $r_0$
and $A(0)$, as it is shown in table \ref{tab2}.

\begin{table}[htp]
\centering
\begin{center}
\begin{tabular}{|c|c|c|}
\hline
$h$  & $r_0$ & $C(R,t,t')$ \\
\hline
$h=0$ & $r_0>-A(0)$ & $C_{\infty}e^{-(r_0+A(0))(t+t')}$ \\
\hline
$h=0$ & $r_0<-A(0)$ & $C_{\infty}+\ e^{2(r_0+A(0))t}$\\
\hline
$h>0$ & & $C_{\infty}+e^{-\frac{1}{2}(\lambda+A(0))(t+t')}$\\
\hline
\end{tabular}
\end{center}
\caption{Long time solutions for $C(R,t,t')$ and $C(R,t)$ at zero temperature for FC protocols.}
\label{tab2}
\end{table}

The fact that auto-correlations decay to
zero or non-zero constants according to the values of $r_0$ and $A(0)$
can be understood in terms of the stability of the component
$\hat{\phi}(k=0)$. For $r_0>-A(0)$ the $k=0$ mode is stable and the
system evolves to a non-magnetized state. When
$r_0<-A(0)$ the instability of the zero mode drives the system to a
non-zero magnetization.

For $h>0$, the decay rate depends on a parameter
$\lambda$ that encloses the dependency on $h$
(see Eq.(\ref{eq:lambda})). The point $\lambda(h_c)=0$ defines the
critical field. Above this value of $h$ the decay rate depends only on
$A(0)$ and measures, essentially, how fast the system relaxes
 to the asymptotic ferromagnetic state.

%*****************************************************************************************
\section{Numerical simulations}
\label{ns}

With the aim to test the predictions of the Hartree approximation we
solved numerically the dynamical equations (\ref{eqts}) for the particular case of
a two dimensional
model of magnetic film with competing exchange and dipolar interactions in
the limit of strong perpendicular anisotropy, in which case the interactions
are perfectly isotropic \cite{garel82}. The continuum nature of the
model and the isotropy of interactions pose some serious challenges to the accuracy
and interpretation of numerical solutions. We devote this section to discuss our
strategy in order to obtain meaningful results. 
We start the section with a discussion of
the spectrum of fluctuations of the model, how it is affected by finite size
and discretization of space and time, that are inevitable when solving differential equations numerically,
and propose some strategies to minimize these effects. After that, we describe the actual
numerical scheme used to solve the equations. Finally, the numerical results are
discussed and compared with the analytic ones from previous sections.

%************************************************
\subsection{The spectrum of fluctuations}
\label{fsa}

We discretize the system in a square mesh of $L\times L$ sites.
 In the discretized reciprocal space, the function $f(t)$ defined in (\ref{ft}) for the
continuum system takes the form:
\begin{equation}
f(t)=\frac{1}{L^2}\sum_{\vec{k}} e^{-2A(k)t},
\label{ftd}
\end{equation}
where 
\begin{equation}
\vec{k}=\left(\frac{2\pi}{a L}\ i,\frac{2\pi}{a L}\ j\right) \ \ \ \ \ \ \ \ \  i,j=-\frac{L}{2}+1,\ldots,\frac{L}{2}, 
\label{kfa}
\end{equation}
$a$ is the linear size of the mesh and $A(k)$ is given by (\ref{AK}). 
From (\ref{ftd}) we should note
that only for $L\gg1$ and a sufficiently small value of the mesh $a$
the discrete version of $f(t)$ can be approximated by the continuum
one.

At this point it is worth to note that, even in the continuum
formulation, the functionality $f(t)\sim1/t^{1/2}$ is only valid for
$t\gg\tau_c$, in which the function under the integral (\ref{ft}) is
sharp enough (see Appendix \ref{Appendix-A} and \cite{mulet07}). 
Moreover, for sufficiently long times, the
discrete nature of $A(k)$ provokes deviations of $f(t)$ from the
expected power law behavior. The characteristic time in which this
deviation happens should be of the order of the inverse of the
difference between $A(k_0)$ and $A(k)$, with $k$ a nearest
neighbor vector of $k_0$.

\begin{figure}[!htb]
\includegraphics[width=5cm,height=7cm,angle=-90]{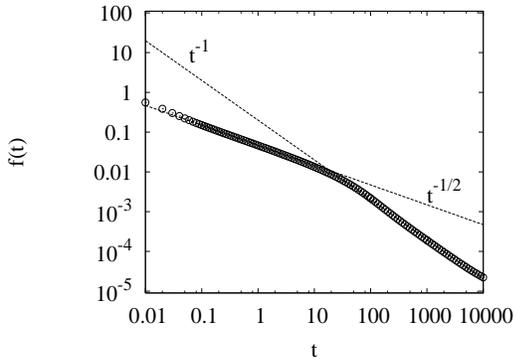}
\caption{Behavior of $f(t)$ for dipolar frustrated ferromagnetic model of size $L=1200$,
 using the $\hat{J}(k)$ given by the Ewald sums. For sufficiently large sizes a crossover 
is observed in the power law exponent from $-1/2$ to $-1$.}
\label{ewfts_ls}
\end{figure}

Up to now we have not specified the nature of the competing interaction
$ J(\vec x,\vec x')$ in (\ref{hamilt}). In ultrathin ferromagnetic
films with strong perpendicular anisotropy, the competing interactions
are the short range exchange (represented in the continuum model by the
square gradient term in (\ref{hamilt})) and the long ranged dipolar
interaction. In the limit of strong perpendicular anisotropy, the
dipolar moments point (except at domain walls) out of the plane of the sample,
 and the 
magnetization can be modeled by a continuum scalar field $\phi(\vec x)$.
In this limit the dipolar term takes the form
$J(\vec x,\vec x')\propto 1/|\vec x - \vec x'|^3$. Note that in this
limit the interactions are perfectly isotropic and so will be the
spectrum of Gaussian fluctuations.
The long range nature of dipolar interactions is commonly taken into
account
by means of the so-called Ewald Summation Technique \cite{lucas06}.
Unfortunately, the use of the $\hat{J}(k)$ given by the Ewald sums
provokes a growing anisotropy in the fluctuation spectrum as the wave
vector $k$ approaches the value $\pi/a$. This makes the minima of the
fluctuation spectrum to remain in a finite number of directions. In
this way, after certain time, the parabolic nature of the isolated
minima provokes a crossover in the exponent of the power law in $f(t)$
from $-1/2$ to a dimensional dependent value $-d/2$, as it is shown in
figure \ref{ewfts_ls} for $d=2$. This new exponent coincides with the
one obtained for the purely ferromagnetic system \cite{NeBr1990,GoLu2002}, where
$f(t)$ is dominated by the single minimum of the fluctuation spectrum
at $k=0$. 

In general, this long time behavior must appear in any system
characterized by several isolated minima in the fluctuation
spectrum. In this sense, it is important to be aware of this subtle
anisotropic effect introduced by the Ewald technique.

In order to avoid these undesired effects, a possibility is to replace
$\hat{J}(k)$ by its expansion for small $k$ up to second order, as has
been already made in literature \cite{Ja2004}. However, a more
justified approach in the context of magnetic thin films is to
consider the two-dimensional Fourier transform of the kernel of the
actual dipolar interaction $1/r^3$. That is
\begin{equation}
\hat{J}(k)=-2\pi k+\frac{2\pi}{\alpha} F\left[ -\frac{1}{2},\left( \frac{1}{2},1\right) ,-\frac{1}{4} k^2 \alpha^2\right] 
\label{jkhyp}
\end{equation}
where $F$ is the confluent hypergeometric function and $\alpha$ is a short distance cutoff.
\begin{figure}[!htb]
\includegraphics[width=5cm,height=7cm,angle=-90]{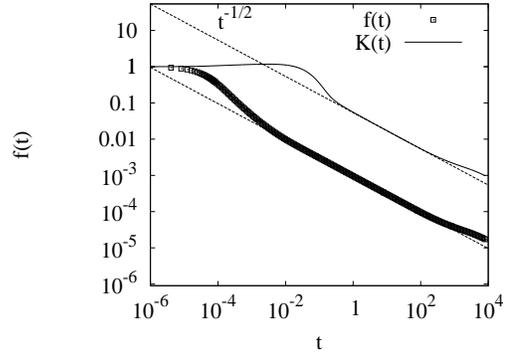}
\caption{Behavior of $f(t)$ for a system of $L=1960$ using the $\hat{J}(k)$ given by expression (\ref{jkhyp}).  
The behavior of $K(t)$ for $h=0$ is also shown. The deviation from power law for large times is a
finite size effect.}
\label{hgft}
\end{figure}

Figure \ref{hgft} shows the behaviors of $f(t)$ and $K(t)$ using the
$\hat{J}(k)$ given by (\ref{jkhyp}).  The range of times in which
$f(t)$ follows a power law with the predicted exponent is determined
only by $\tau_c$ and an upper bound dependent on the size of the
system.

%*****************************************
\subsection{Numerical scheme}
\label{nsch}

In order to build a numerical scheme it is useful to express
(\ref{eqts}) in a dimensionless form. A natural choice for the
characteristic length is the wave length corresponding to the minimum
of the spectrum of fluctuations $\lambda_0=2\pi/k_0$. Taking
$J(\vec{x},\vec{x}')$ as the dipolar interaction with a short distance
cut-off $\alpha$ described in the last subsection, this characteristic
length is given by $\lambda_0=2\delta\left(1+\pi\alpha/2\delta
\right)$.  Defining also a characteristic time
$t_0=\lambda_0^2/\Gamma$ and a characteristic field value
$\phi_s=\sqrt{r/u}$, we can define new reduced variables as
\begin{eqnarray}
\frac{\vec{x}}{\lambda_0}\rightarrow\vec{x},\ \ \frac{t}{t_0}\rightarrow t,\ \ 
\frac{\delta}{\lambda_0}\rightarrow\delta,\ \ \frac{\alpha}{\lambda_0}\rightarrow\alpha, \ \
\frac{\phi}{\phi_s}\rightarrow\phi.
\label{dimtransf}
\end{eqnarray}
Equation (\ref{eqts}) can be rewritten as:
\begin{eqnarray}
\nonumber
\frac{\partial \phi(\vec{x},t)}{\partial t}&=& \nabla^2 \phi(\vec{x},t) 
-\frac{1}{\delta} \int d^2 \vec{x}' J(|\vec{x}-\vec{x}'|, \alpha) \phi(\vec{x}',t)\\
&+& \beta [\phi(\vec{x},t)-\phi^3(\vec{x},t)]
+ h +  \sqrt{2 T} \,\eta(\vec{x},t)
\label{eqtsD}
\end{eqnarray}
where we have defined the dimensionless variables
\begin{eqnarray}
\frac{h \lambda_0^2}{\phi_s}\rightarrow h,\ \ \ \ \lambda_0^2 r \rightarrow\beta,
\ \ \ \ \frac{T}{\phi_s^2} \rightarrow T,
\end{eqnarray}
and the thermal noise has been normalized in such a way that
$\left\langle\eta(\vec{x},t)\eta(\vec{x}',t')\right\rangle=\delta(t-t')
\delta(\vec{x}-\vec{x}')$. So, we can write in the reciprocal space
\begin{eqnarray}
%\nonumber
\frac{\partial \hat{\phi}(\vec{k},t)}{\partial t}&=& 
-\left[ k^2+\frac{1}{\delta}\hat{J}(k)-\beta\right]\ \hat{\phi}(\vec{k},t) \nonumber \\
&+& \left[h - \beta\phi^3(\vec{x},t) + \sqrt{2 T} \eta(\vec{x},t) \right]_{\vec{k}}^F 
\label{dfdtad}
\end{eqnarray}
where $]_{\vec{k}}^F$ means the $\vec{k}$ component of the corresponding Fourier transform.

We discretize the space with a mesh size $a=1/M$.  Since $\lambda_0$
is our characteristic length, this implies that the wave-length of the
modulated structures consists in approximately $M$ sites of the
simulation system. Now, if we want the system to be able to show $N$
modulated structures, we must simulate a square lattice with a linear
size of $L=MN$ sites.  Thus, we construct a scheme in which $M$ and
$N$ are fixed at a large enough value to ensure realistic simulations
for the continuum model.  In this sense, both quantities play a
well-defined role. Increasing $M$ is associated with an improvement of
the domain wall dynamics as the magnetic structures can extend over
several mesh sites. This results in a better definition of the time
$\tau_c$ in which $f(t)$ reaches its long times behavior. On the other
hand, increasing $N$ allows us to see a larger fraction of the
(ideally infinite) system and is related to the growth of the maximum
time for which $f(t)$ can still be approximated to its long time
functionality.

In order to further simplify the spatially discretized dimensionless
equation we now repeat the transformations in equation
(\ref{dimtransf}) for both space and time variables using $a$ in place
of the characteristic length ($\vec{k}a\rightarrow\vec{k}$ and
$t/a^2\rightarrow t$), and redefine the strength of the local
potential as $a^2 \beta \rightarrow \beta$. Finally we consider a
standard first-order semi-implicit spectral integration scheme, with a
time step $dt$, to construct the following recurrence relation:
\begin{eqnarray}
\nonumber
\hat{\phi}(\vec{k},t+dt)&=&\frac{\hat{\phi}(\vec{k},t)}{1+k^2dt}\left[ 1 + dt(\beta - a \hat{J}(\vec{k}) /\delta )\right] \\ \nonumber
&+&\frac{ dt \, \left[ - \beta\phi^3(\vec{x},t)+  a^2 h + \sqrt{\frac{2 T}{dt}}\eta(\vec{x},t)
\right]_{\vec{k}}^F}{ 1 + k^2 dt }
\end{eqnarray}
where the noise term $\eta$ is a random Gaussian number with unit
variance. Here we are taking advantage of the isotropic form of the
laplacian in the inverse space, where the adimensional wave vector has
the form $\vec{k}=(2\pi\,i/L, 2\pi\,j/L)$.

%**********************************************
\subsubsection{Choosing simulation parameters}

Simulations were carried out for a system of $M=35$ and $N=56$, that
is $L=1960$, with the parameters $\alpha=0.04$, $\beta=0.265$. From
the chosen form of the dipolar interaction we have $\delta$ fixed as
$\delta=(1-\pi\alpha)/2$. The value of $M$ was chosen to be large
enough as to guaranty the convergence of $\tau_c$ to the continuum
system value.  To have a good estimate of the simulation time needed
to reach the asymptotic regime of the model, we first solved
numerically equation (\ref{K}) (see figure \ref{hgft}).  One can see
that the time at which the asymptotic functionality of $K(t)$ sets in
has a lower bound different from that of $f(t)$, this bound is
dependent on the whole set of parameters and is greater than $\tau_c$.

Comparing our numerical scheme with equation (\ref{eqts}) it is
possible to express the parameters $r_0$, $A(0)$ and $g$ used in
analytical calculations as
\begin{eqnarray}
\nonumber
r_0&=&{\cal A}(k_0)\\ \nonumber
A(0)&=&{\cal A}(0)-{\cal A}(k_0)\\ 
g&=&3\beta
\label{calig}
\end{eqnarray} 
where ${\cal A}(k)$ is the numerical fluctuation spectrum.  In this
way the critical field can be written as
\begin{equation}
h_c=\frac{1}{a^2}\sqrt{\frac{-{\cal A}(k_0)}{3\beta}}\  [{\cal A}(0)-{\cal A}(k_0)] 
\label{hcred}
\end{equation}
that, under our particular set of parameters takes the value $h_c=4.53$.

%********************************************************************************
\subsection{Simulation results}
\label{sr}

In our simulations, the system is initialized in a disordered
(homogeneous) configuration corresponding to the ZFC (FC)
 experiment. Suddenly, a subcritical temperature is
fixed and the system is let to evolve a time $t_w$, simulating a
quench. Then,
autocorrelations are computed as function of the time $\Delta t$
elapsed after $t_w$. 
%From these definitions we can do the associations
%$t'=t_w$ and $t=t_w+\Delta t$.

\subsubsection{Field Cooling}

We present first, the dynamical behavior of the system after a quench
from a fully magnetized state for $T=0$ and $h<h_c$. In particular, we
focus our attention of the differences between the cases $r_0>-A(0)$
and $r_0<-A(0)$ at $h=0$ (see table \ref{tab2}).  Tuning the value of $\beta$, 
and starting from the same initial conditions, one can study both cases. 

\begin{figure}[!htb]
\includegraphics[height=5cm]{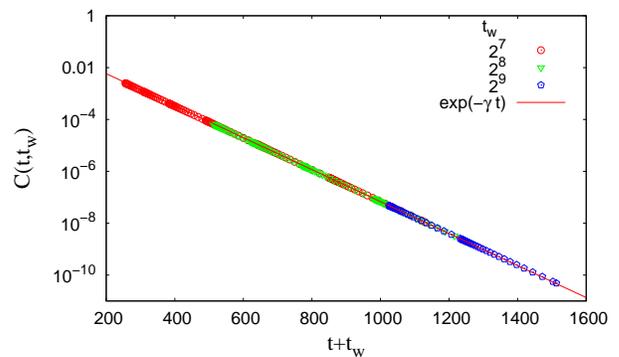}
\caption{Self-correlations for $r_0>-A(0)$  in the $T=0$ and $h=0$ ferromagnetic quench.}
\label{ferroct_ma}
\end{figure}

\begin{figure}[!htb]
\includegraphics[height=5cm]{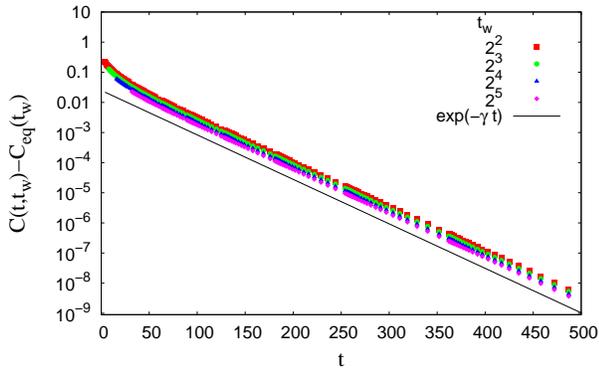}
\caption{Self-correlations for $r_0<-A(0)$ in the $T=0$ and $h=0$ ferromagnetic quench.}
\label{ferroct_mi}
\end{figure}

In figures \ref{ferroct_ma} and \ref{ferroct_mi}  we show autocorrelations for  $r_0>-A(0)$ 
and $r_0<-A(0)$ respectively. In both cases, the numerical estimations fit very well with 
the asymptotic predictions in Table \ref{tab2}. In addition also the exponential factors 
characterizing the decay rates ($r_o+A(0)$) of these correlations  are consistent with the 
analytical predictions. The difference between the theoretical values for the parameters 
used and the numerical estimation obtained fitting the numerical data is smaller than 5 percent.

At zero temperature, as we already mentioned, a system evolving from
homogeneous ferromagnetic initial conditions is incapable to break the spatial
homogeneity.  In this context, the fact that autocorrelations decay to
zero or non-zero constants according to the values of $r_0$ and $A(0)$
can be understood in terms of the stability of the component
$\hat{\phi}(k=0)$. For $r_0>-A(0)$ one has from (\ref{calig}) that
${\cal A}(0)>0$. This in turn implies that the $k=0$ mode is unstable and the
system evolves to a non-magnetized state. On the contrary, for
$r_0<-A(0)$ the stability condition ${\cal A}(0)<0$ drives the system to a
non-zero magnetization.

\subsubsection{Zero Field Cooling}

\begin{figure}[!htb]
\includegraphics[width=2cm,height=2cm]{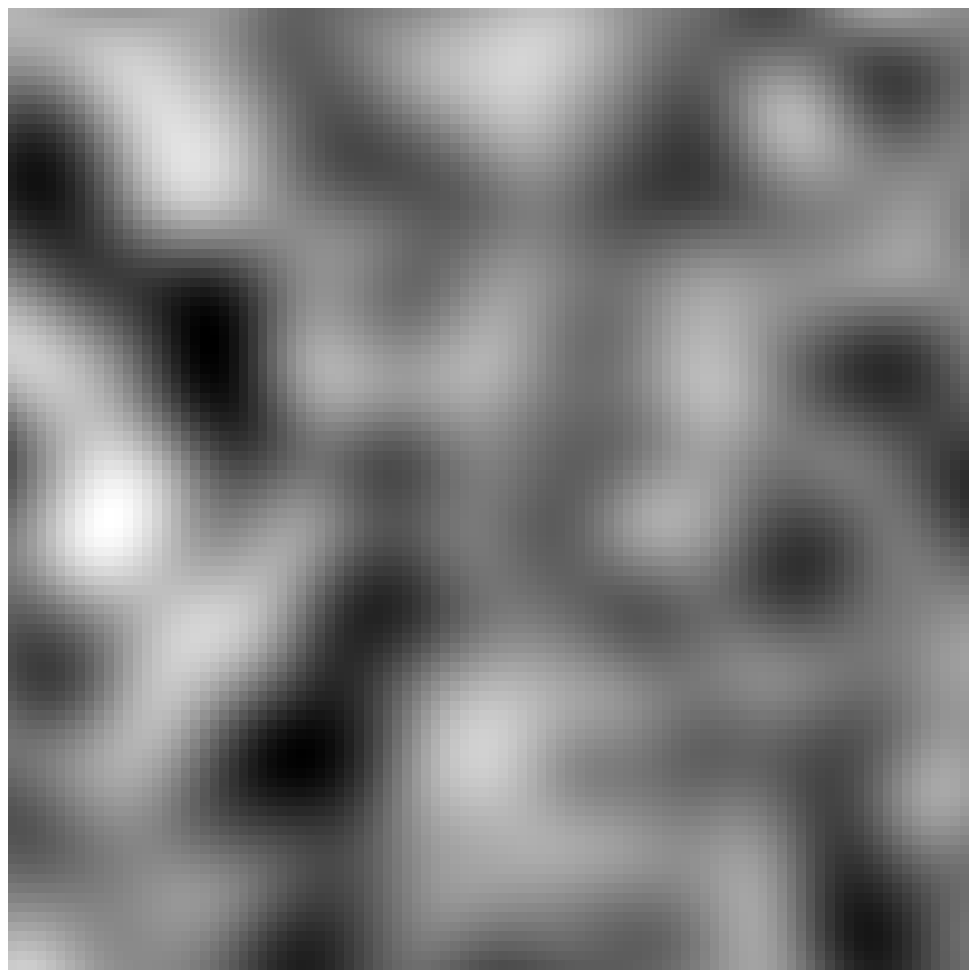}
\includegraphics[width=2cm,height=2cm]{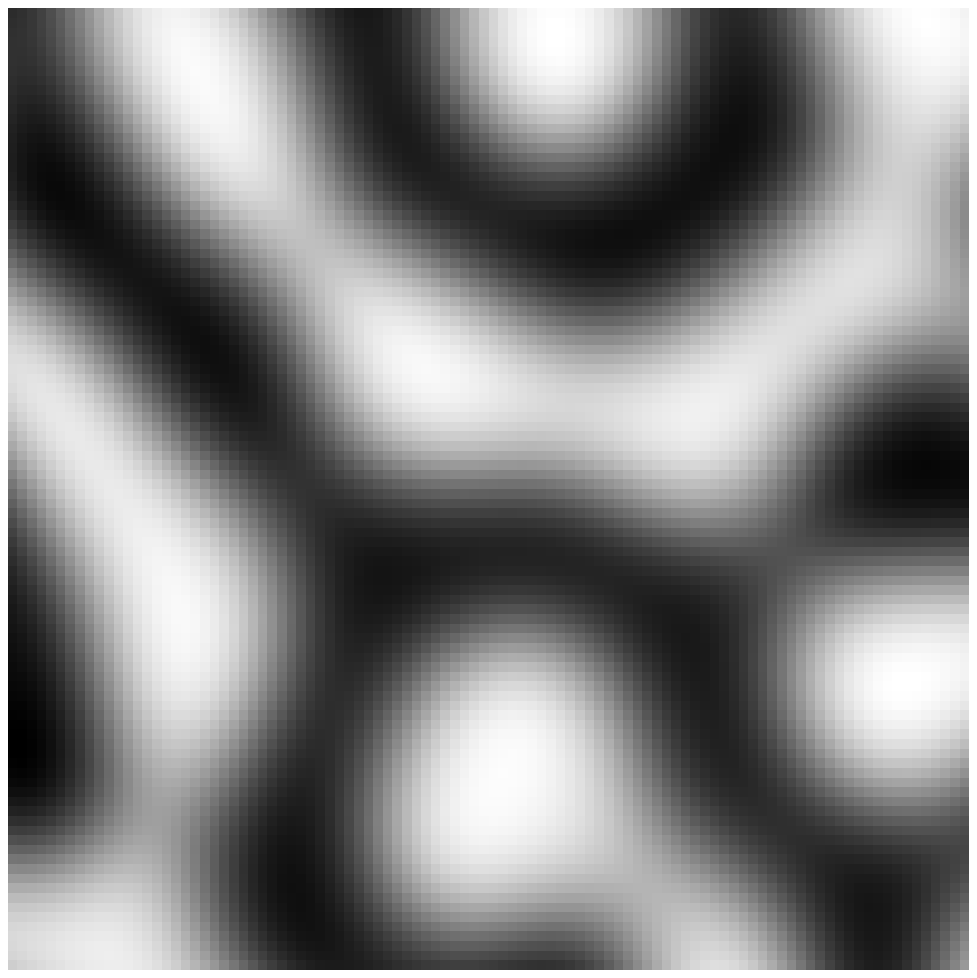}
\includegraphics[width=2cm,height=2cm]{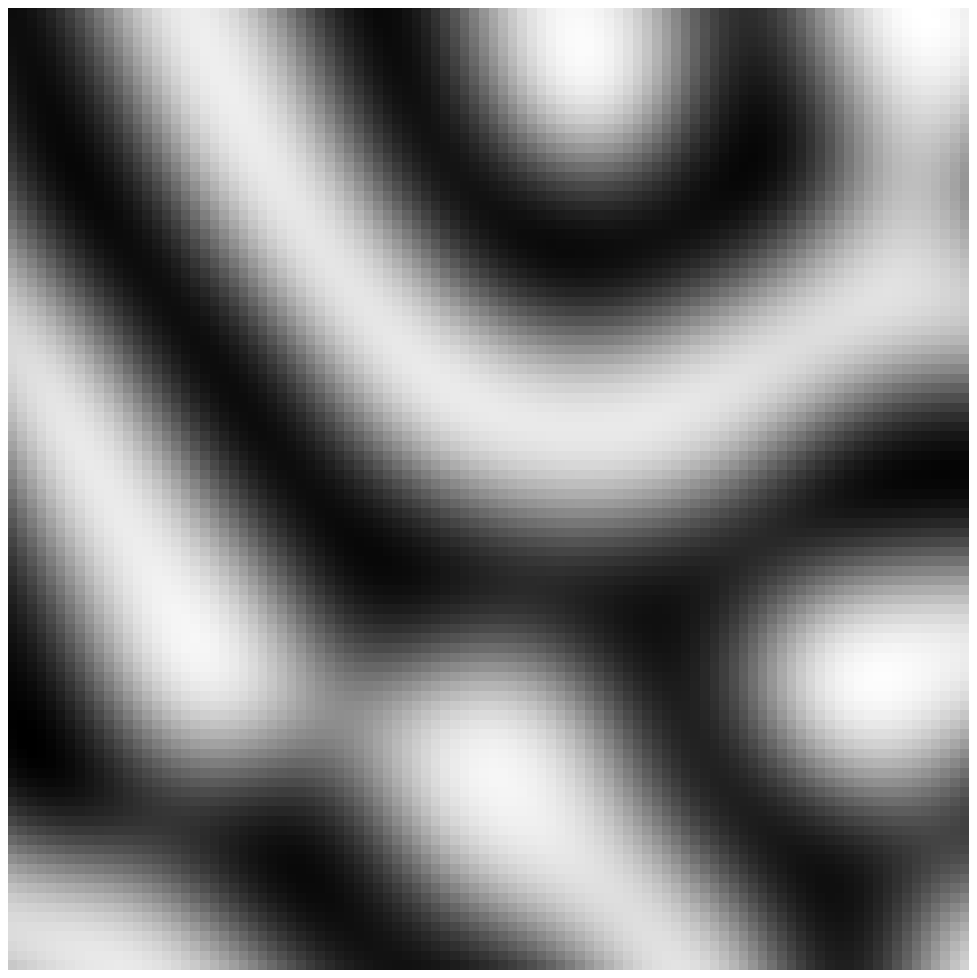}
\includegraphics[width=2cm,height=2cm]{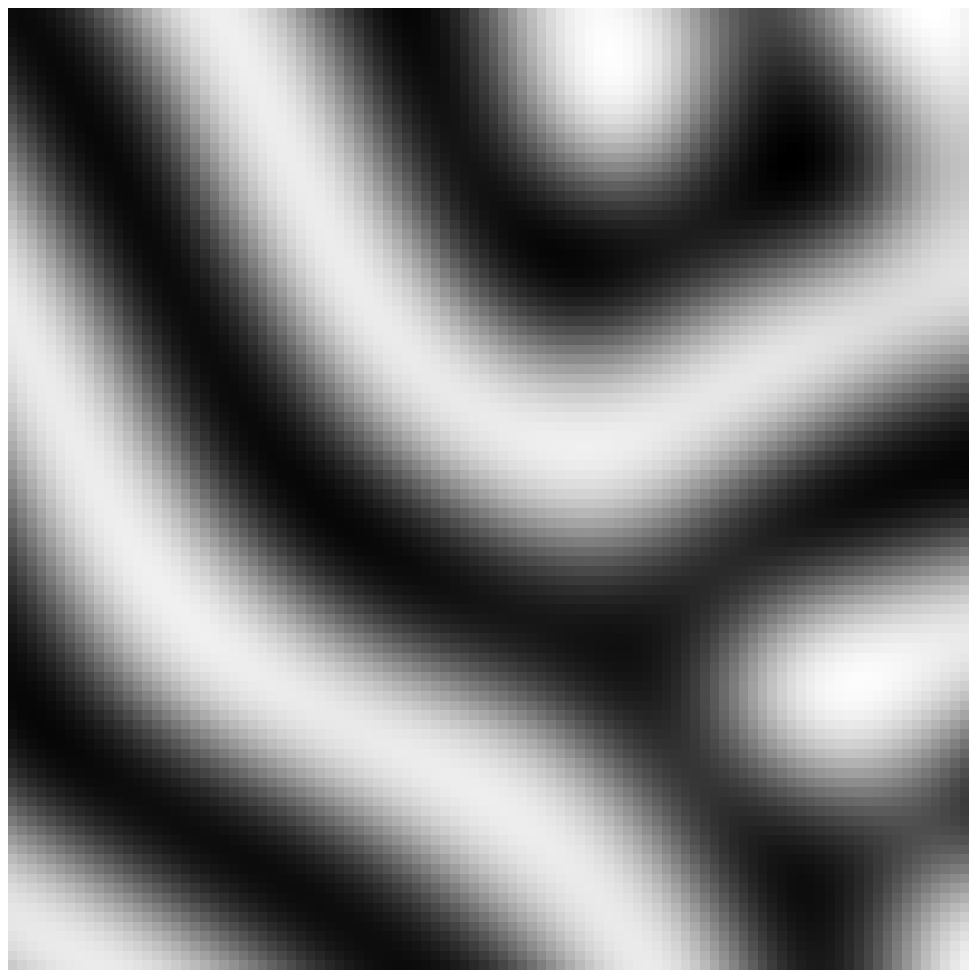}\\
\includegraphics[width=2cm,height=2cm]{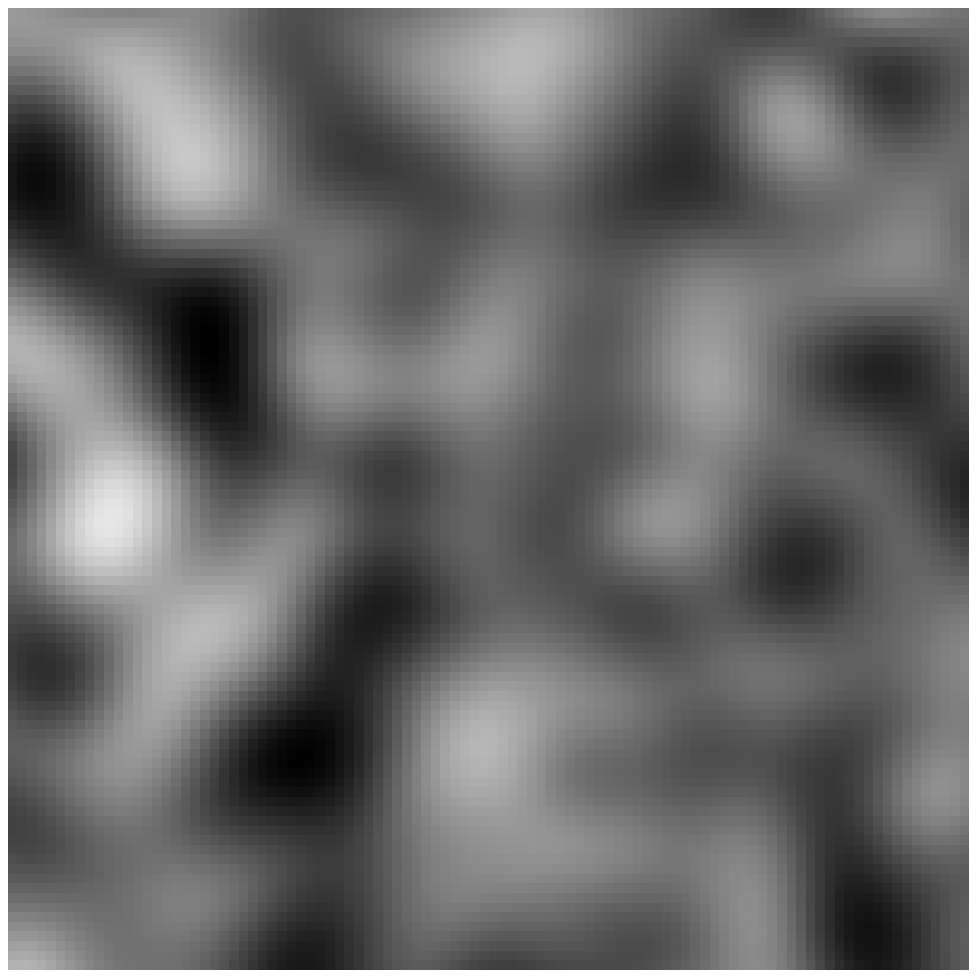}
\includegraphics[width=2cm,height=2cm]{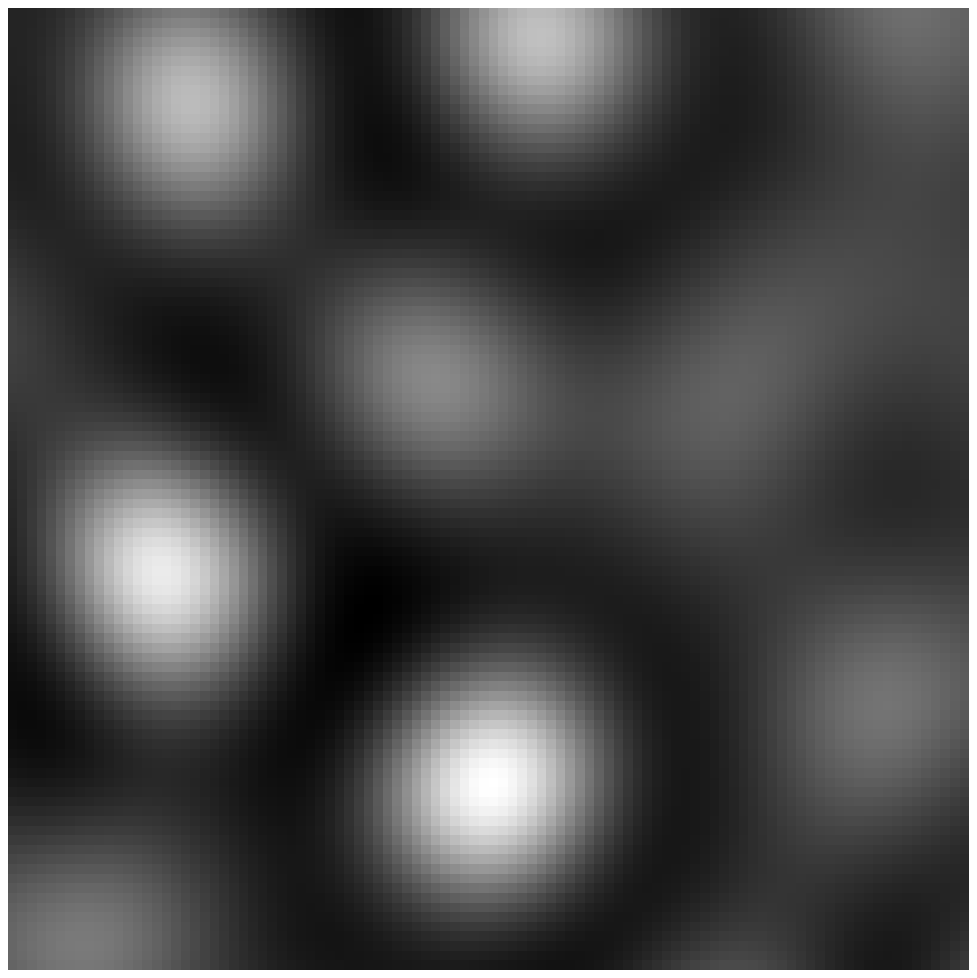}
\includegraphics[width=2cm,height=2cm]{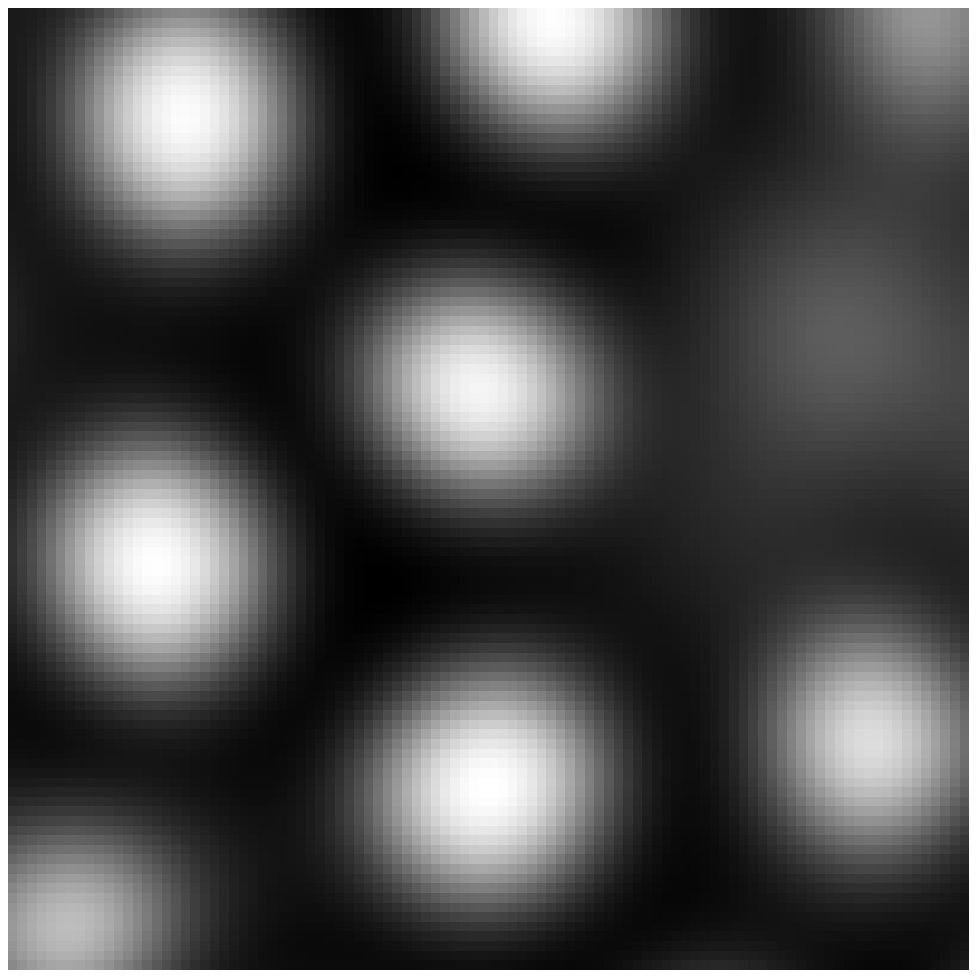}
\includegraphics[width=2cm,height=2cm]{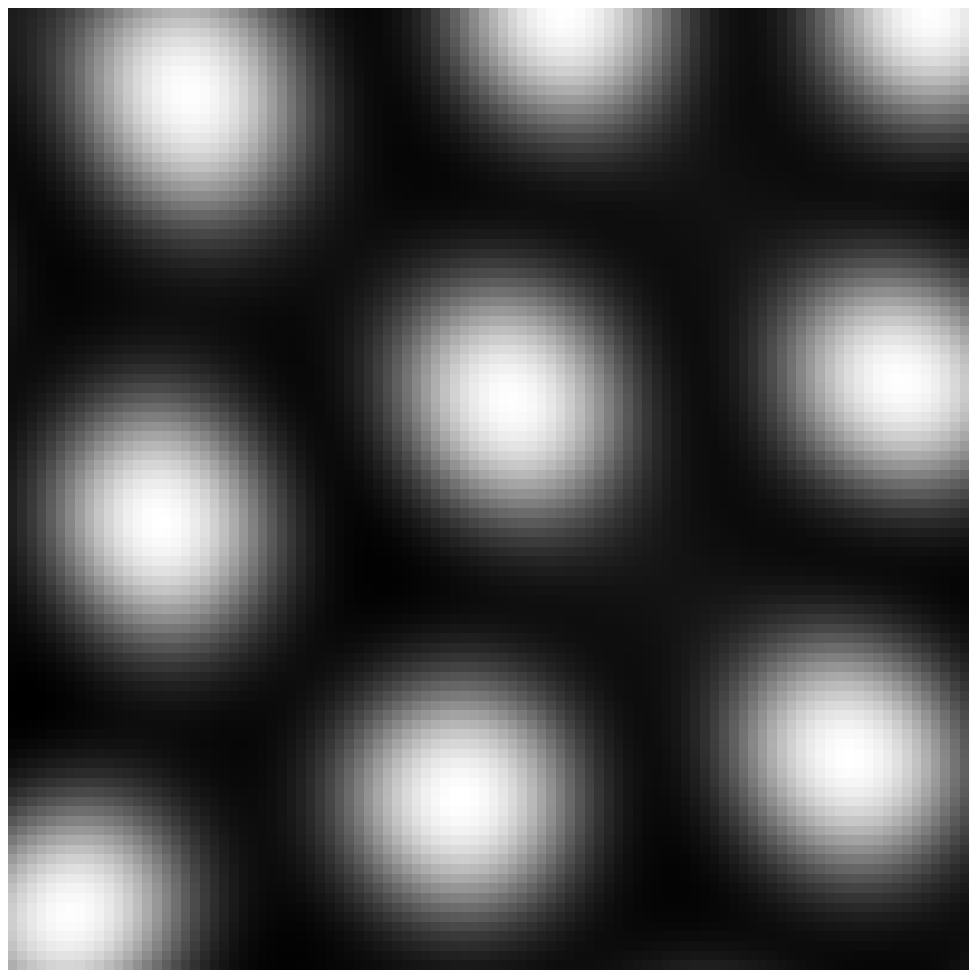}\\
\includegraphics[width=2cm,height=2cm]{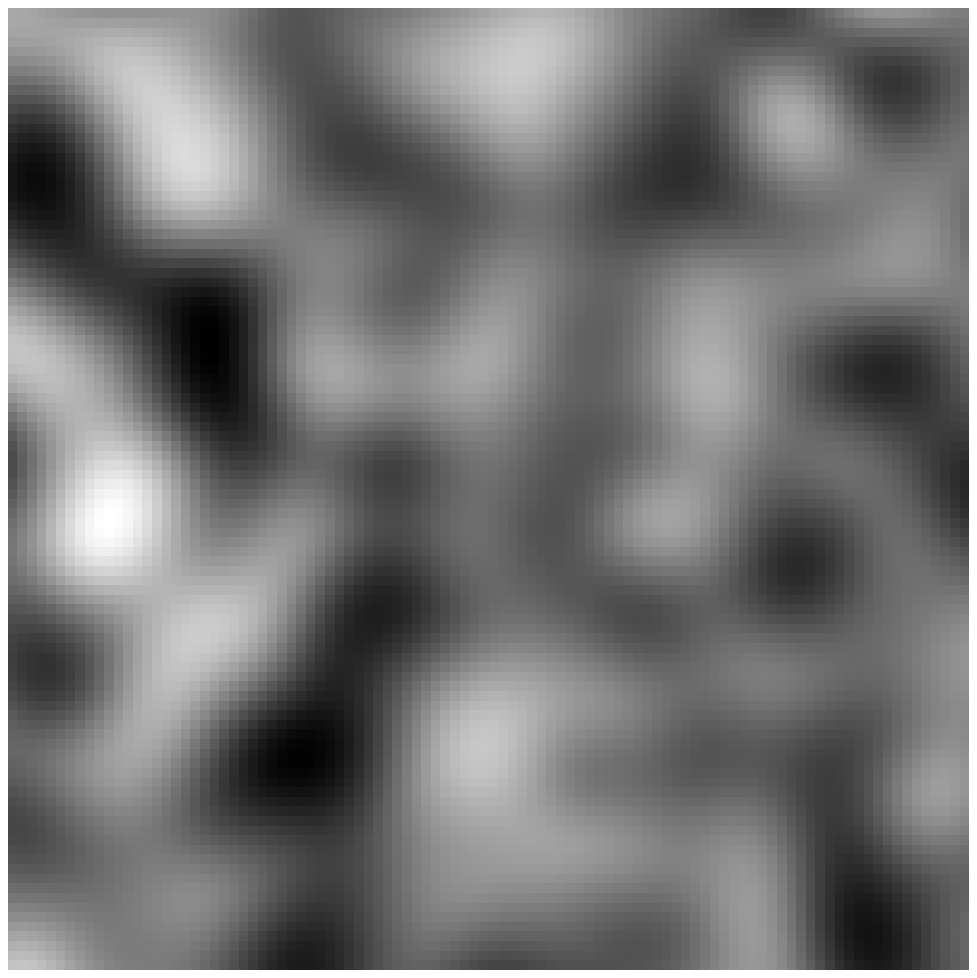}
\includegraphics[width=2cm,height=2cm]{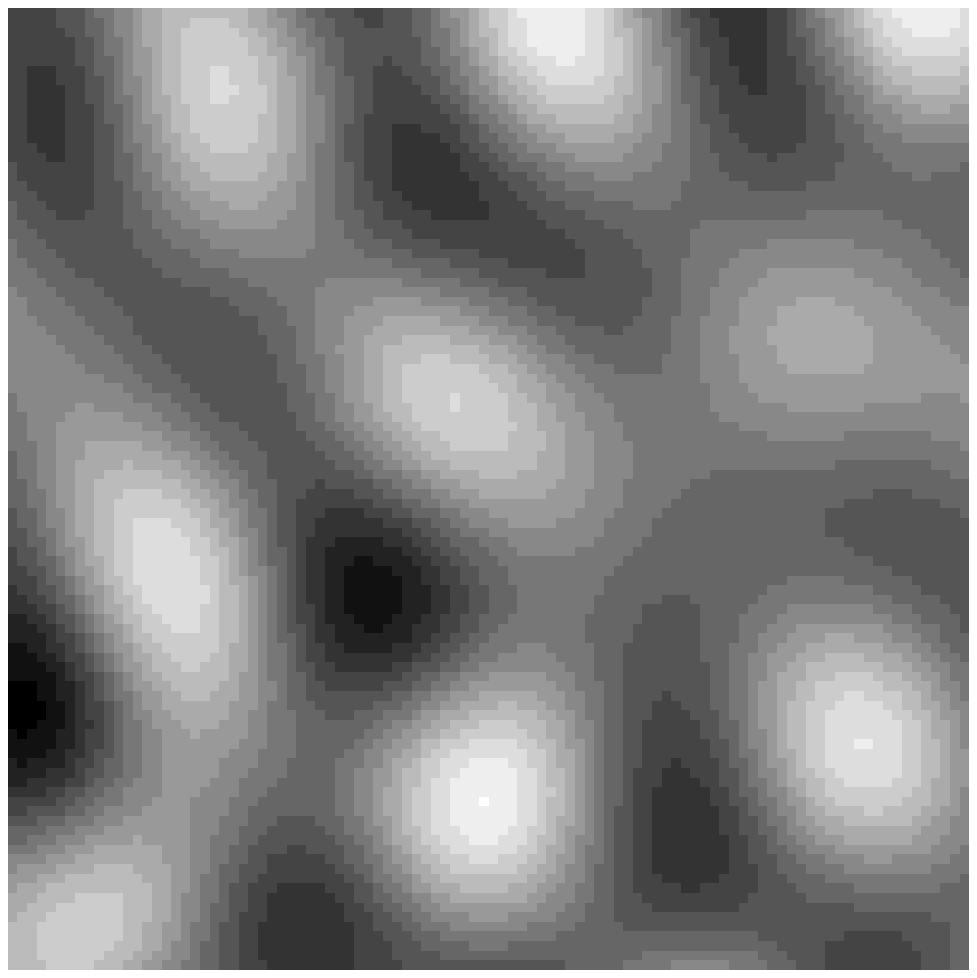}
\includegraphics[width=2cm,height=2cm]{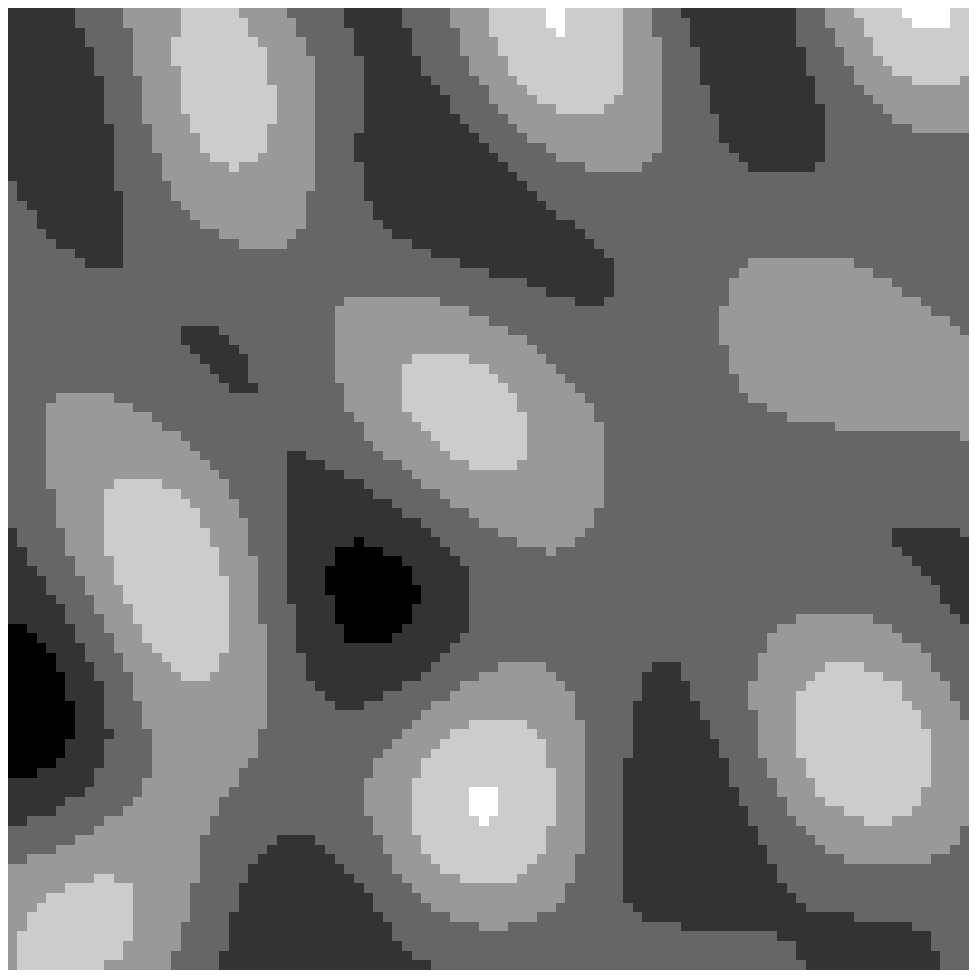}
\includegraphics[width=2cm,height=2cm]{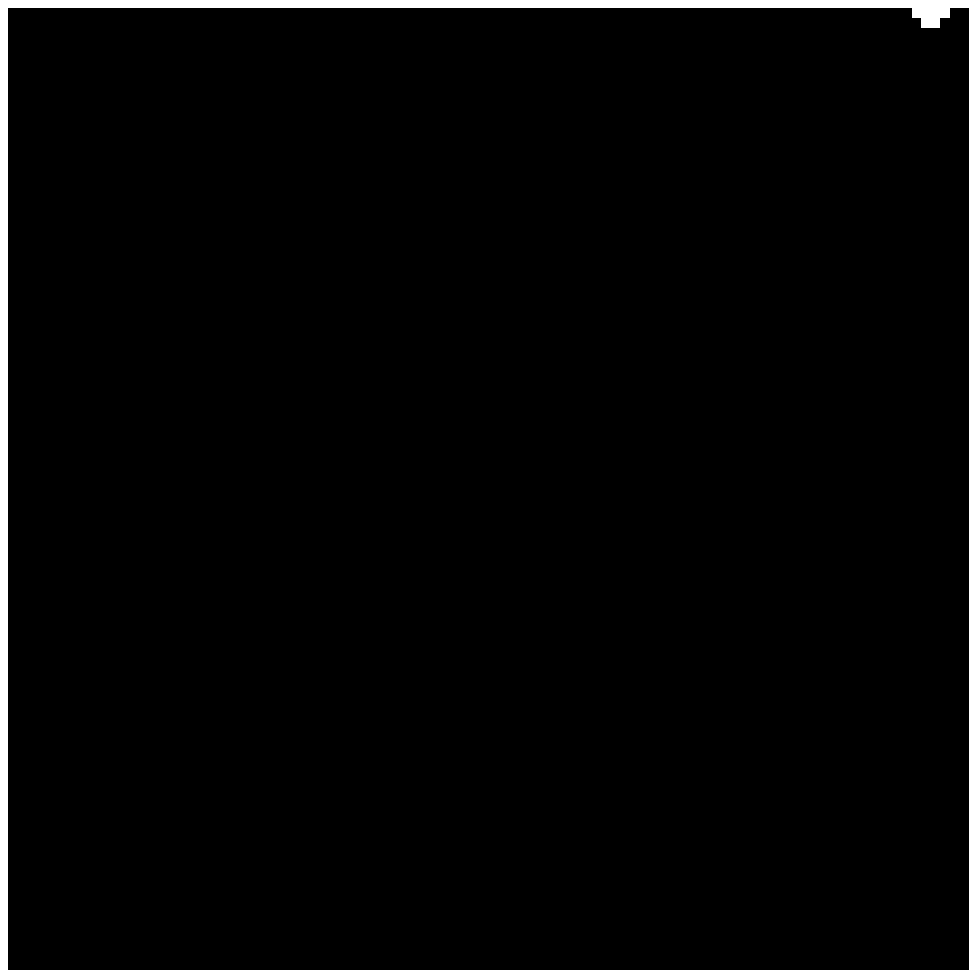}
\caption{Evolution of the magnetization patterns of a $100\times100$
portion of the $1960\times1960$ system for three different fields after a disordered
quench. First row: $h=0$, second row: $h=4.0<h_c$ and third row:
$h=4.7>h_c$. Time grows from left to right.}
\label{fotos}
\end{figure}

The effects of the predicted critical field (\ref{hcr}) can be
seen in figure \ref{fotos}.

There, three dynamical sequences are shown after a disordered quench
for field values above and below the critical one. It can be seen that
final configurations evolve, correspondingly, to modulated or
homogeneous states as expected from our results (see Table \ref{tab1})
and also predicted by thermodynamic equilibrium studies~\cite{garel82}.

%In figure \ref{hghcTg0a} the exponential relaxation for $T=0$ and $h>h_c$ 
%predicted by the analytical results is confirmed by our simulations.

\begin{figure}[!htb]
\includegraphics[height=5cm]{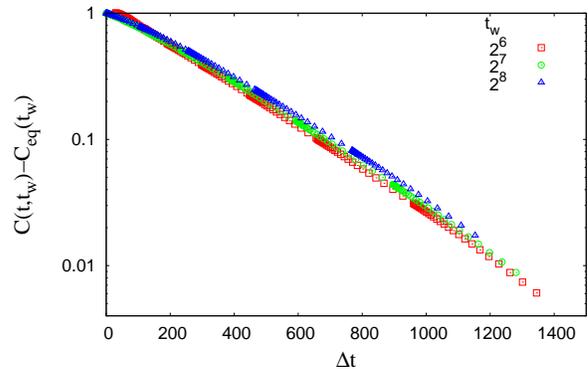}
\caption{Autocorrelations in the ZFC case for $T=0$ and $h>h_c$ and
several waiting times. An asymptotic exponential relaxation is observed
in agreement with analytic results in table \ref{tab1}.}
\label{hghcTg0a}
\end{figure}

Autocorrelations for $T=0$ and $h>h_c$ decay very rapidly to a
plateau which is dependent on $t_w$.  In figure \ref{hghcTg0a}
the quantity $C_{eq}(t_w)$ has been subtracted.  It is seen that, as time 
grows, the exponential decays become evident, in agreement with 
analytical results in table \ref{tab1}.

%\begin{figure}[!htb]
%\includegraphics[height=5cm]{./collapsem35n56T.1.eps}
%\caption{Interrupted aging for $T=0.1$ and $h=0$ in the disordered quench.}
%\label{hghcTg0b}
%\end{figure}

%Figure \ref{hghcTg0b} shows the autocorrelations for $T=0.1$ where one
%can see clear evidence of interrupted aging. While there is a $t_w$
%dependence in the relaxation for the smaller waiting times, a 
%$t_w$-independent decay in $\Delta t$ is gradually approached as
%$t_w$ increases, and eventually a time translational invariant
%regime sets in.

%The testing of the more interesting aging region predicted for
%disordered quench at $T=0$ and $h<h_c$ reveals, anyway, some
%disagreement with Hartree calculations. As we have explained before,
%the expected scenario for it is a simple aging with an exponent that
%tends to $1/4$ for any $t_w$ in the long times limit. In fact, what
%simulations reveals is a rather sub-aging scenario.
\begin{figure}[!htb]
\includegraphics[height=5cm]{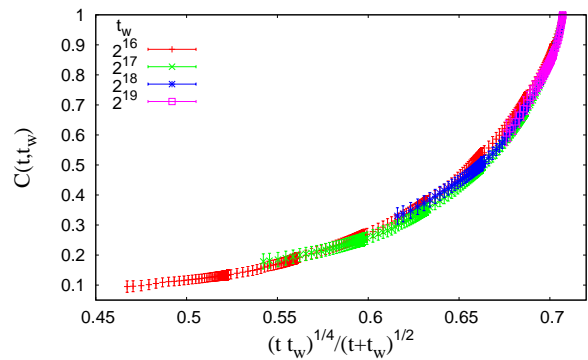}
\caption{Collapse of the autocorrelations for the ZFC case,
at $T=0$ and $h=0$, using 12 samples.}
\label{clpT0h0}
\end{figure}

For $h=0$, figure \ref{clpT0h0} shows a reasonable agreement with the predicted 
collapse of the two-time autocorrelations (see Table \ref{tab1} and \cite{mulet07}) 
in the asymptotic regime, attained for waiting times greater than $5\times10^5$ time
steps.

\begin{figure}[!htb]
\includegraphics[height=5cm]{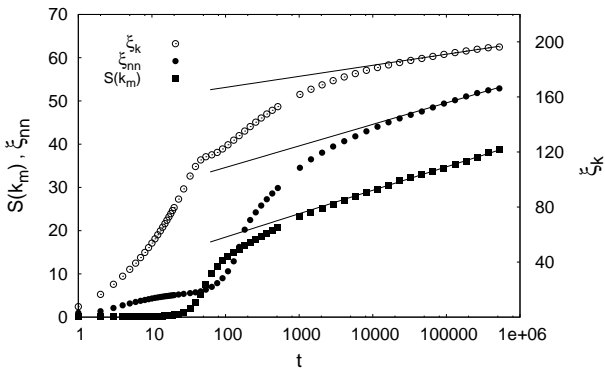}
\caption{Growing order in the disordered quench dynamics at $T=0$ and
$h=0$ through the characteristic orientational and positional length
scales ($\xi_{nn}(t)$ and $\xi_{k}(t)$), and the peak of the azimuthal
average of the structure factor ($\hat{S}(k_m)$), shown in a
log-linear scale. Error bars are smaller than point size and the solid
lines corresponds to logarithmic fits and are guidelines for the eye
only.}
\label{lengths}
\end{figure}

Finally, an important result concerns the growth law of domains at low
temperatures, as discussed in the introduction. Our analytical results
of eq. (\ref{growthlaw}) predict $\xi_k(t)\simeq t^{1/2}$, where $\xi_k(t)$
is a positional correlation length. In figure \ref{lengths} we show 
the time dependence of the azimuthally averaged structure factor, the
positional correlation length and the orientational correlation length,
defined in \cite{lucas06}. At long times all quantities seem to follow
a logarithmic growth, at variance with the power law with $z=2$ found
analytically. This slower regime may be due to pinning of topological
defects, as observed also in a different model by Gomez et al. \cite{GoVaVe2006},
which is not captured by the Hartree approximation. More work is need
to fully elucidate the origin of the logarithmic relaxation observed in
figure \ref{lengths}. 

%*****************************************************************************************
\section{Conclusions}
\label{conc}

We studied the Langevin dynamics of a model with competing isotropic interactions
at different scales at $T=0$ in the presence of an external field. Analytic results were 
found in the self-consistent field
(Hartree) approximation for two-time autocorrelation functions and also for
spatial correlations. Two different dynamical protocols (initial conditions)
were considered: zero field cooled and field cooled. Analytical results were
confronted with numerical simulations of a prototypical system consisting of
a ferromagnetic interaction frustrated with long range dipolar interactions
in two dimensions. A detailed numerical scheme was developed and discussed.

Both analytically and numerically we found a critical field
$h_c$ which divides the long time dynamic behavior of the model in two regions:
for $h\leq h_c$, the long time dynamics strongly depends on the
particular initial condition, while for $h > h_c$ the initial
conditions are irrelevant. The regime $h\leq h_c$ corresponds to the
appearance of modulated structures at low temperatures (stripes or bubbles), 
while for $h > h_c$ the system is a paramagnet in an external field. 

We obtained explicit expressions for the autocorrelations and
spatial correlation functions, both for the disordered and ferromagnetic
quenches, considering all possible values of temperature and external
applied field. In the FC case relaxations are essentially exponential.
 In the ZFC case
the system presents a slow coarsening of modulated domains. This
coarsening process reflects in an aging behavior of the two-time
correlation functions. The analytical prediction for the aging scaling
is in reasonable agreement with the numerical results. The results
for the spatial correlations predict a growth law for the modulated
domains with a power law $\xi \simeq t^{1/2}$. This is not verified numerically,
but a logarithmic growth $\xi \simeq \ln{t}$ at long times is observed instead.
This slow logarithmic behavior may be consequence of the very slow 
drift of topological defects, which are not taken into account by the
self-consistent field approximation. 

Finally, it is important to remark that although the Hartree approximation
does not consider the influence of topological defects in the dynamical
behavior, our numerical results at $T=0$ compare quite well with the
predictions of the theory. Surely topological defects are very important
in the long time dynamics of the system and must be included in a
complete theory of dynamics of modulated systems. Nevertheless, after a
quench at zero temperature the dynamics of defects is extremely slow,
such that our approximate theory can still give valuable information
on the relaxation in a time scale where defects can be considered as
effectively ``quenched''.

\begin{acknowledgments}
We gratefully acknowledge partial financial support from the Abdus
Salam ICTP through grant {\em Net-61, La\-tin\-a\-me\-ri\-can Network
on Slow Dynamics in Complex Systems}. Calculation facilities kindly
offered by the Bioinformatic's Group of the Center of Molecular
Immunology in Cuba were instrumental to this work. 
L.~N. and D. A. S.~would like to thank partial financial support from CNPq,
Brazil.
\end{acknowledgments}

\appendix

%***************************************************************************
\section{Solution for $K(t)$ at $T=0$}{\label{Appendix-A}}

In this case equation (\ref{K}) for $K(t)$ takes the form:
\begin{eqnarray}
\nonumber
\frac{dK(t)}{dt}&=&2r_0K(t)+2gV(t)\\ \nonumber
&+&4gh\ K(t)^{1/2}\phi_0 e^{-A(0)t}\ \times \\ \nonumber
&\ &\ \times \int_0^td\tau\ K(\tau)^{1/2}e^{-A(0)(t-\tau)}\\ 
&+&2gh^2\left[\int_0^t d\tau\ K(\tau)^{1/2}e^{-A(0)(t-\tau)} \right]^2
\label{KT0}
\end{eqnarray}
Here a hypothesis is done as a very important step of the
calculation that will allow us to linearize equation (\ref{KT0}),
namely:
\begin{eqnarray}
\nonumber
\int_0^t K^{1/2}(t')e^{-A(0)(t-t')}dt' &=& \zeta K^{1/2}(t) \\
A(0)t &\gg& 1,
\label{hyp}
\end{eqnarray}
and its consistency will be proved in the solution process, since the parameter $\zeta$ is
determined self-consistently using the hypothesis itself. By means of
(\ref{hyp}), equation (\ref{KT0}) reads
\begin{eqnarray}
\nonumber
\frac{dK(t)}{dt}&=&2r_0K(t)+2gV(t) \\
&+&4gh\ \phi_0 e^{-A(0)t}\ \zeta K(t)+2gh^2\zeta^2K(t).
\end{eqnarray}
The term $4gh\ \phi_0 e^{-A(0)t}\ \zeta
K(t)$ is unimportant in the long time limit, because it decays
faster than others in the equation, given that $A(0)>0$. Then, in the
long time limit, we have to solve the following ordinary linear differential
equation:
\begin{equation}
\frac{d K(t)}{dt}=(2r_0+2gh^2\zeta^2)K(t)+2gV(t).
\end{equation}

The general solution of this equation is
\begin{eqnarray}
\nonumber
K(t)&=&K(0)e^{(2r_0+2gh^2\zeta^2)t}\\
&&+2g\int_0^t d\tau \ V(\tau)e^{(2r_0+2gh^2\zeta^2)(t-\tau)}.
\label{solK}
\end{eqnarray}
Different kind of solutions in the
long time limit are possible according to the sign of 
$2r_0+2gh^2\zeta^2$.

If $2r_0+2gh^2\zeta^2>0$, then:
\begin{eqnarray}
\nonumber
K(t)&=&\left(K(0)+2g\int_0^\infty d\tau\ V(\tau)e^{-(2r_0+2gh^2\zeta^2)\tau}\right) \times \\
&&\times \  e^{(2r_0+2gh^2\zeta^2)t}.
\end{eqnarray}
The integral converges because $V(t)$ is a bounded, decreasing
function of $t$. In this case we obtain a general solution
independently of the particular function $V(t)$.

If $2r_0+2gh^2\zeta^2=0$:
\begin{equation}
K(t)=K(0)+2g\int_0^\tau d\tau\ V(\tau).
\label{hc}
\end{equation}
This solution is less general than the previous one. It is valid while
the term given by the higher order correction to the hypothesis
(\ref{hyp}) is irrelevant in comparison with $V(t)$.

If $2r_0+2gh^2\zeta^2<0$, like before, there is not a general
solution. The long time solution is given by the term decaying slower
in expression (\ref{solK}).

It is possible to obtain a critical field as the lowest applied field $h_c$ for which
$K(t)$ is an exponential. To do this we suppose that 
$2r_0+2gh^2\zeta^2>0$. Then:
\begin{equation}
K(t)=Ce^{(2r_0+2gh^2\zeta^2)t},
\end{equation}
and with condition (\ref{hyp}) we obtain the following equation for the parameter $\zeta$:
\begin{equation}
\zeta=\frac{1}{r_0+gh^2\zeta^2+A(0)}.
\end{equation}
The limit case for which we have an exponential solution is when
$2r_0+2gh_c^2\zeta^2=0$, applying this condition to the previous
equation we get $\zeta_c=1/A(0)$. Then, the critical field is given by:
\begin{equation}
h_c=\sqrt{-\frac{r_0}{g}}A(0).
\label{hcr}
\end{equation}
This field defines a zone in the parameter space ($h>h_c$) in which
the long time behavior of the system is universal, in the sense that
it is independent of initial conditions.

%***************************************************************************
\subsection{$h\leq h_c$}{\label{skt}}

We considered separately quenches from the disordered and ferromagnetic 
initial conditions.

%******************************************
\subsubsection{ZFC, disordered case }

With conditions (\ref{gcond}) appropriately taken, the equation
(\ref{KT0}) for the ZFC system reads:
\begin{eqnarray}
\nonumber
\frac{dK(t)}{dt}&=&2r_0K(t)+2g\Delta f(t)\\
&+&2gh^2\left[\int_0^t d\tau\ K(\tau)^{1/2}e^{-A(0)(t-\tau)} \right]^2
\label{Kdis}
\end{eqnarray}
where, 
in the long time limit, $f(t)$ is given by~\cite{mulet07}:
\begin{equation}
f(t)=\alpha \ t^{-\frac{1}{2}},
\label{ftat12}
\end{equation}
with $\alpha=\frac{k_0^{d-1}}{(2\pi)^d}\frac{2 \pi^{\frac{d}{2}}
}{\Gamma(\frac{d}{2})}\sqrt{\frac{\pi}{A_2}}$.

We need to solve this equation considering the applied field to be
lower than or equal to the critical one. In order to solve
equation (\ref{Kdis}), we use hypothesis (\ref{hyp}) obtaining the
general solution:
\begin{eqnarray}
\nonumber
K(t)&=&K(0)e^{(2r_0+2gh^2\zeta^2)t}\\
&+&2g\alpha \Delta \int_0^t  \frac{d\tau}{\tau^{\frac{1}{2}}} e^{(2r_0+2gh^2\zeta^2)(t-\tau)}.
\end{eqnarray}
Considering now $2r_0+2gh^2\zeta^2<0$ (that is $h<h_c$) we get, in
the long time limit:
\begin{equation}
\nonumber
K(t)=\frac{-2g\alpha \Delta}{(2r_0+2gh^2\zeta^2)t^{\frac{1}{2}}} .
\end{equation}
Applying condition (\ref{hyp}) we obtain
$\zeta=\frac{1}{A(0)}$, which allows us to rewrite $K(t)$ as:
\begin{equation}
K(t)=\frac{\alpha \Delta A^2(0)}{(h_c^2-h^2)t^{\frac{1}{2}}} .
\label{Kdh<hc}
\end{equation}

When $h=h_c$ ,
higher order
corrections to the hypothesis (\ref{hyp}) 
ive:
\begin{equation}
K(t)=4g\alpha \Delta \frac{A(0)}{A(0)-2r_0} t^{\frac{1}{2}}.
\label{Kdhc}
\end{equation}

%**********************************************
\subsubsection{FC, ferromagnetic case }

Using the initial conditions (\ref{gcond}) equation
(\ref{KT0}) becomes:
\begin{eqnarray}
\frac{dK(t)}{dt}&=&2r_0K(t)+2g\phi_0^2\ e^{-2A(0)t} \nonumber \\ 
&+&4gh\phi_0\ e^{-A(0)t}\int_0^t d\tau\ K^{1/2}(\tau)e^{-A(0)(t-\tau)} \nonumber \\ 
&+& 2gh^2\left[\int_0^t d\tau\ K(\tau)^{1/2}e^{-A(0)(t-\tau)} \right]^2
\label{Kfer}
\end{eqnarray}
and, as before, using hypothesis (\ref{hyp}) it is possible to arrive
to the following general solution in the long time limit:
\begin{equation}
K(t)=K(0)e^{\lambda t}+2g\phi_0^2 \int_0^t d\tau \ e^{-2A(0)\tau} e^{\lambda(t-\tau)},
\end{equation}
where $\lambda=2r_0+2gh^2\zeta^2$.

The analysis of the above solution when $h<h_c$ reveals different possibilities
according to the different values of the parameters
involved. In the long time limit we obtain:

\underline{$h=0$}
\begin{displaymath}
K(t)=\left\{ \begin{array}{ll}
(1+\frac{g\phi_0^2}{r_0+A(0)})\ e^{2r_0t} &  \textrm{for $r_0>-A(0)$} \\ \\
2g\phi_0^2\ te^{-2A(0)t} &  \textrm{for $r_0=-A(0)$} \\  \\
-\frac{g\phi_0^2}{r_0+A(0)}\ e^{-2A(0)t} &  \textrm{for $r_0<-A(0)$} \\ \\
\end{array} \right.
\end{displaymath}

\underline{$h\neq0$}
\begin{equation}
K(t)=Ce^{\lambda t}.
\end{equation}
$\lambda$ is given by the solution of the equation
\begin{equation}
\lambda=2r_0+\frac{2gh^2}{\left(A(0)+\frac{\lambda}{2}\right)^2},
\label{eq:lambda}
\end{equation}
where hypothesis (\ref{hyp}) has been used. Note
that for all $h\leq h_c$ then
$\lambda\leq0$, being $\lambda(h_c)=0$. 

\bibliography{paper_dinamica_v1.4c}

\end{document}